# Auto-resolving the atomic structure at van der Waals interfaces using a generative model

Wenqiang Huang[1,2,3,†], Yucheng Jin[4,5,6,†], Zhemin Li[7,†], Lin Yao[8]*, Yun Chen[2], Zheng Luo[2], Shen Zhou[7], Jinguo Lin[9], Feng Liu[9], Zhifeng Gao[8], Jun Cheng[4,5,6], Linfeng Zhang[8,10], Fangping Ouyang[3]*, Jin Zhang[1]* & Shanshan Wang[1,2,11]*

[1] School of Advanced Materials, Peking University Shenzhen Graduate School, Shenzhen 518055, China

[2] Department of Materials Science and Engineering, College of Aerospace Science and Engineering, National University of Defense Technology, Changsha 410000, China

[3] School of Physics, Central South University, Changsha 410083, China

[4] State Key Laboratory of Physical Chemistry of Solid Surfaces, Collaborative Innovation Center Materials (iChEM), College of Chemistry and Chemical Engineering, Xiamen University, Xiamen 361005, China

[5] Innovation Laboratory for Sciences and Technologies of Energy Materials of Fujian Province (IKKEM), Xiamen 361005, China

[6] Institute of Artificial Intelligence, Xiamen University, Xiamen 361005, China

[7] College of Science, National University of Defense Technology, Changsha 410000, China

[8] DP Technology, Beijing 100080, China

[9] State Key Laboratory of Nonlinear Mechanics, Institute of Mechanics, Chinese Academy of Sciences, Beijing 100190, China

[10] AI for Science Institute, Beijing 100080, China

[11] Guangdong Provincial Key Laboratory of Nano-Micro Materials Research, Peking University Shenzhen Graduate School, Shenzhen 518055, China

[†]These authors contributed equally.

*Corresponding authors: L.Y. (yaol@dp.tech); F.O. (oyfp@csu.edu.cn); J.Z. (jinzhang@pku.edu.cn); S.W. (wangshanshan08@nudt.edu.cn)

**The high-resolution visualization of atomic structures is significant for understanding the relationship between the microscopic configurations and macroscopic properties of materials. However, a rapid, accurate, and robust approach to automatically resolve complex patterns in atomic-resolution microscopy remains difficult to implement. Here, we present a Trident strategy-enhanced disentangled representation learning method (a generative model), which utilizes a few unlabelled experimental images with abundant low-cost simulated images to generate a large corpus of annotated simulation data that closely resembles experimental results, producing a high-quality large-volume training dataset. A structural inference model is then trained via a residual neural**

network which can directly deduce the interlayer slip and rotation of diversified and complicated stacking patterns at van der Waals (vdW) interfaces with picometer-scale accuracy across various materials (e.g. $MoS_2$, $WS_2$, $ReS_2$, $ReSe_2$, and 1T'-$MoTe_2$) with different layer numbers (bilayer and trilayers), demonstrating robustness to defects, imaging quality, and surface contaminations. The framework can also identify pattern transition interfaces, quantify subtle motif variations, and discriminate moiré patterns that are difficult to distinguish in frequency domains. Finally, the high-throughput processing ability of our method provides insights into a vdW epitaxy mode where various thermodynamically favorable slip stackings can coexist.

**Introduction**

Two-dimensional (2D) van der Waals (vdW) materials have profoundly expanded the design space of artificial solids by manipulating the relative rotation and slip between adjacent atomic planes. The interlayer twist generates a moiré superlattice with long periodic potential that triggers exotic physical properties like superconductivity[1,2], ferroelectricity[3,4], and nontrivial topological states[5,6]. The relative sliding between layers alternates atomic registries which enables to tailor material's electrical[7–10], magnetic[11–13], and catalytic properties[14]. Scanning transmission electron microscopy (STEM) can provide sub-angstrom scale configuration information by raster scanning across the sample surface atom-by-atom and becomes a powerful tool to unclose the structure-property relation of various nanostructures including 2D vdW materials[15–17], thus laying the foundation for rational structural design and accurate performance prediction by theoretical calculations. However, the analysis of STEM images still predominantly relies on human experts, which suffers from long time consumption, inferior labeling accuracy, individual identification bias, poor tolerance for image imperfections (e.g. low image signal-to-noise ratio (SNR)[18], surface contaminations[19], etc.), limited analytical capability for complex patterns, and low-throughput hindered discovery of statistically grounded clues.

Machine learning (ML) algorithms bring opportunities for automatic classification, identification, and feature extraction of atomic-resolution images, which, in principle, can be divided into unsupervised and supervised strategies. Unsupervised learning is capable of uncovering inherent patterns and relationships of unlabelled data by techniques of clustering, association, and dimensionality reduction without upfront human intervention and has been employed to classify point defects[20–22] and extract structural motifs[23]. However, the output of unsupervised learning often suffers from subjectivity and inferior interpretability due to a lack of predefined target variables, which restricts the tasks that the method can handle, makes objective evaluation of the model performance difficult, and commonly requires domain knowledge from human experts to achieve comprehensible and insightful discoveries. Supervised learning is another well-established approach that has been utilized to localize atomic columns[21,24–29], identify vacancies and dopants[24,30–32], segment polymorphs[30,33,34], categorize crystal structures[34–37], assign molecular chirality[38] and stacking orders[39], etc. This method has explicit performance evaluation (measured by errors between real and predicted values through the loss function), high prediction accuracy, generalizability to unseen data, and flexibility for a wide range of tasks.

However, supervised learning requires a large corpus of annotated data for model training, which comes from either manual labelling of experimental images[27,38,39] or simulated data generated by software with ground truth[25,26,29,30,33,34,40]. The former has high training data quality but suffers from laborious labeling work, poor annotation accuracy, and scarcity of experimental images. The latter can satisfy data sufficiency at a low cost. However, the quality of the simulated data is severely inferior to experimental images due to prominent discrepancies in the visual style, which is determined by various factors including detector noise, scanning distortion, lens aberrations, surface contamination, etc. The difficulty in simultaneously achieving high quality and large volumes of the training dataset makes the current application of supervised learning in atomic-resolution image analysis restricted to simple questions, such as classifying limited types of microstructures with obvious pattern disparity, with inference accuracy highly sensitive to the image presentation state (e.g. SNR[18], defects[17,41,42] and contamination[19], image drift[43] and astigmatism[44], etc.). Can we extend the ability of supervised learning algorithms from "identifying" discrete and finite microstructures, which is a simple classification question (e.g. differentiating atomic defects from pristine lattices that show obvious and fixed structural disparity) to "solving" complex patterns with almost continuous variations and subtle disparities, which is a regression question in a higher difficulty level (e.g. outputting accurate interlayer slip vectors and twist angles corresponding to different stacking patterns whose atomic position differences are only on the picometer scale)? Clark et al. pioneered the introduction of a cycle generative adversarial network (CycleGAN) to augment simulated STEM images with realistic spatial frequency information, which has demonstrated feasibility in identifying atomic defects[32]. However, the imperfection of CycleGAN in strictly maintaining image contents after style transfer limits the scope of the scientific problems that can be handled by this algorithm (Supplementary Fig. 1). Therefore, it is highly desirable to develop a fast, accurate, and robust supervised learning framework that can automatically accomplish complex structural analysis tasks.

In this paper, we develop a Trident strategy-enhanced disentangled representation (DR) learning approach[45], which utilizes a small set of unlabelled experimental STEM images with abundant low-cost simulated images to generate a large annotated training dataset that closely resembles experimental image styles with the simulation image contents strictly maintained after style conversion, thus showing a superior balance between the quality and quantity of training data. A structural inference model is then trained by these high-quality simulated images using a residual neural network, which enables direct output of the interlayer slip and rotation of diversified and complicated stacking patterns in an end-to-end manner with an accuracy of picometer level. Our framework can also identify stacking pattern transition interfaces, quantify subtle motif variations with a high spatial resolution, and discriminate moiré patterns that are difficult to distinguish in the frequency domains. Our model demonstrates robustness to defects, imaging quality, and surface contaminations and can be generalized to various vdW materials (e.g. $MoS_2$, $WS_2$, $ReS_2$,

ReSe$_2$, and 1T'-MoTe$_2$) and different layer numbers (e.g. bilayer and trilayers).

## Results

### Overview of the ML framework

The first step of the framework is to train a generative model via a Disentangled Representation for Image-to-Image Translation (DRIT) algorithm[45] that can produce high-quality STEM simulated images (Fig. 1a). It is realized by combining the structural information (e.g. position, brightness, and size of atoms) from the software-generated low-quality, noise-free simulated images with the visual style from the experimental images. The second step is to define structural descriptors for slip and twisted stackings which enable to represent all potential stacking configurations, followed by the generation of realistic STEM simulated images via the well-trained DRIT model in the first step (Fig. 1b). A large training dataset having precise labels and high stylistic similarity with experimental images is thus achieved, alleviating the problem of data scarcity caused by the high cost of the STEM experiments and the inefficiency of atom-by-atom manual labeling of experimental images. The descriptor of the slip stacking is a slip coordinate ($D_a$, $D_b$) acquired by the decomposition of the slip vector **D** along two in-plane base vector directions of the monolayer unit cell, while for the twisted stacking, an interlayer rotation angle $\theta$ is applied (Supplementary Fig. 2). The third step is to train an end-to-end stacking structure identification model using a ResNet-50 architecture as the backbone of the regression network. The relations between the stacking structural labels, i.e. ($D_a$, $D_b$) and $\theta$, and the realistic STEM simulated images are learned by the two ResNet models, respectively (Fig. 1c), thus enabling the straightforward, accurate, and efficient auto-resolving of the interlayer sliding and twist at the vdW interface from the experimental images.

    The key to the overall workflow is the DRIT model training, which determines whether the abundant, low-cost but also low-quality STEM simulated images can be successfully transformed into the high-quality counterparts with structural information strictly unchanged and visual style greatly resembling the experimental images so that large training data can be obtained for the following supervised learning. Two points require in-depth comprehension. One is the reason for selecting the DRIT algorithm for the style transfer. The other is the modifications required to the basic DRIT model for better task accomplishment.

    For the first point, it is essential to recognize two primary challenges of our visual effect transformation task. First, the training data is severely unbalanced, where the simple, noise-free STEM simulated images can be easily generated in batches by the Computem software but the experimental images from which the realistic visual style can be extracted are scarce. Second, the stacking structure is much more complex than the previously discussed point defects (e.g. vacancies and dopants), displaying obvious configuration diversity and subtle structural discrepancy (at sub-angstrom scale) between different stackings. Fortunately, the DR learning algorithm represented by the DRIT model is proficient in solving these difficulties. It decouples an image into two distinct spaces: a domain-invariant content space capturing shared structural information across the noise-free simulated images and the experimental images, and a domain-specific attribute space independently extracting the visual styles from the noise-free simulated images and the

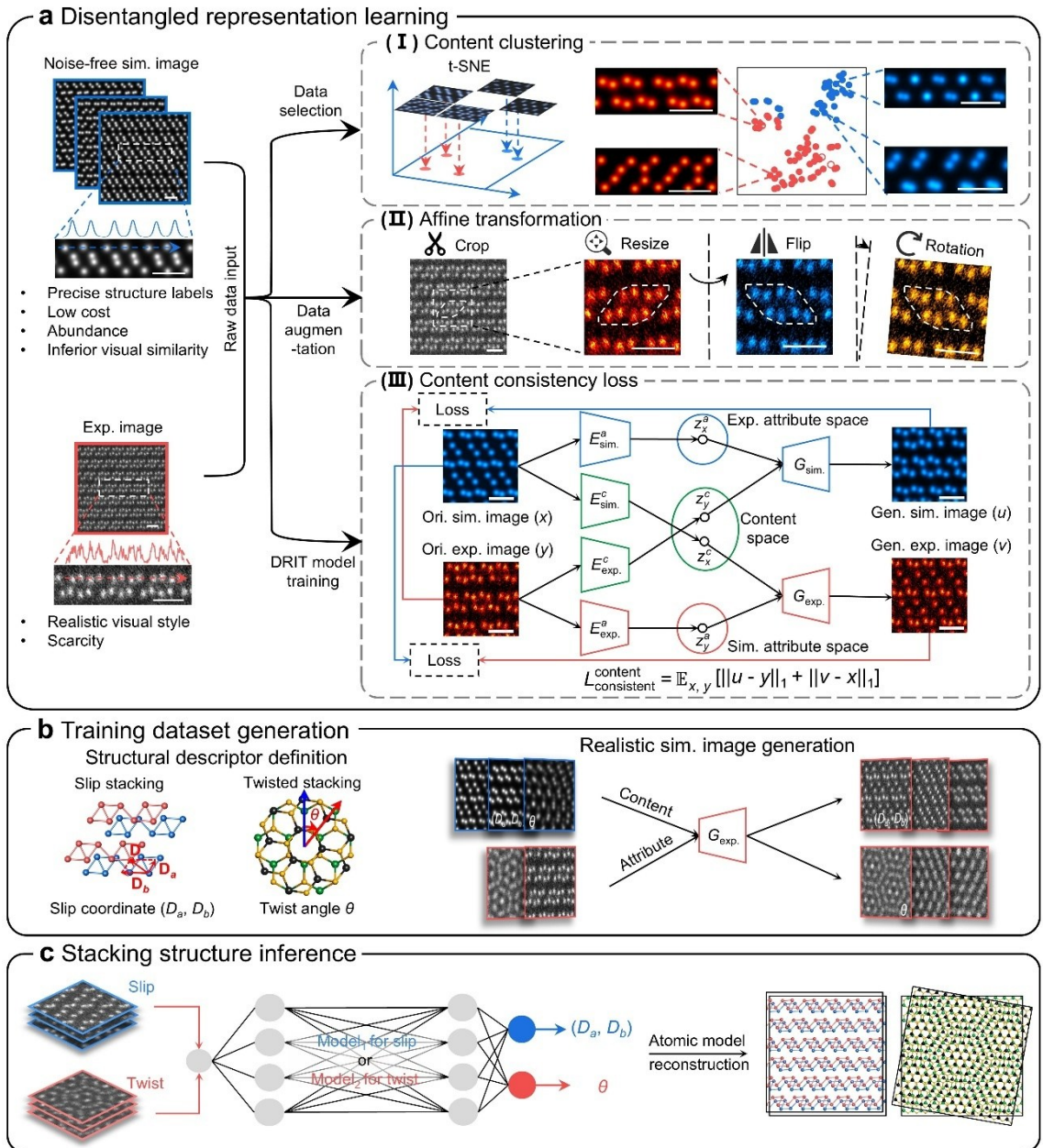

**Fig. 1 Machine learning (ML) workflow. a**, Disentangled representation learning approach for high-quality simulated image generation. A Trident strategy is developed, including image content clustering via t-distributed stochastic neighboring embedding (t-SNE) algorithm (I), affine transformation (II), and addition of a content consistency loss function $L_{\text{consistent}}^{\text{content}}$ (III) to preserve atomic structure during image style transformation. Words of experiment, simulation, original, and generated are abbreviated as exp., sim., ori., and gen., respectively. The blue and red lines above the rectangular sim. and exp. images in the left of (a) are the intensity line profiles taken along the dashed lines, suggesting the attribute discrepancy between the Computem-generated simulated image and the experimental image. The content encoder, attribute encoder, and generator for simulation and experimental domains are denoted as $E_{\text{sim.}}^c$, $E_{\text{sim.}}^a$, $G_{\text{sim.}}$ and $E_{\text{exp.}}^c$, $E_{\text{exp.}}^a$, $G_{\text{exp.}}$, respectively. Scale bars: 0.5 nm. **b**, Training dataset generation involving defining the structural descriptors for slip and twisted stackings and generating high-quality simulated images using the well-trained generative model in (a). ReS$_2$ is applied to represent the slip-stacked bilayer atomic model, matching with the case study in Fig. 3. Red and blue spheres indicate Re atoms in different layers. S atoms are hidden for visual clarity. MoS$_2$ is applied to represent the twist-stacked atomic model, matching with the case study in Fig. 5. Black and green spheres represent Mo atoms in different layers, while yellow spheres represent S atoms. **c**, Establishment of the stacking structure inference models which directly output the slip coordinate $(D_a, D_b)$ and the twist angle $\theta$ for slip- and twist-stacked experimental scanning transmission electron microscopy (STEM) images, respectively, based on which precise atomic models can be automatically constructed. Panel (a)(III) is adapted with permission from ref. 45, Springer Nature.

experimental images. Such algorithm architecture together with three strategies, i.e., weight sharing, a content discriminator, and a cross-cycle consistency loss, ensures effective visual style conversion with the image contents strictly unchanged for unbalanced and unpaired training data involving complex structural information (Supplementary Figs. 3-5). Moreover, DR learning aligns with the human cognitive pattern, exhibiting explicit translation stage by stage, thus improving the interpretability and comprehensibility of the algorithm.

For the second point, despite the superiority of the basic DRIT model in style transformation to some other algorithms such as CycleGAN[46], we developed a "Trident strategy" to further strengthen the reliability and robustness of the model. Three operations were conducted: (i) Data selection. We applied an unsupervised learning method called the t-distributed stochastic neighboring embedding (t-SNE) algorithm to topologically cluster the large numbers of programmatically generated noise-free simulated images and select those that adopt relatively high structural similarity with the experimental images for the subsequent DRIT model training (Fig. 1a(I)) (see Methods). This step is especially significant when the number of the experimental images employed for the model training is minimal (e.g. only three experimental STEM images were used for the bilayer $ReS_2$ slip stacking training task) since it helps the model avoid misinterpreting the image features belonging to the domain-invariant content space (structure information) to the domain-variant attribute space (visual style). (ii) Data augmentation. Affine transformations (e.g. cropping, scaling, rotation, and flipping) which preserve collinearity and distance ratios of points but alter and diversify their absolute coordinates were then applied on both experimental and selected simulated images, thus enhancing the ability of the DRIT algorithm in maintaining the structure information during the image style conversion (Fig. 1a(II)). (iii) Content consistency loss. The experimental and simulated STEM images after data selection and augmentation were finally introduced to the DRIT model for training (Fig 1a(III)), whose architecture involves the content encoders ($E^c_{\text{exp.}}$, $E^c_{\text{sim.}}$) enciphering the atomic structures from both the experimental and simulated images into a mutual content space, the attribute encoders ($E^a_{\text{exp.}}$, $E^a_{\text{sim.}}$) enciphering the visual style from the experimental and the simulated images independently into their respective attribute spaces, and the generators ($G_{\text{exp.}}$, $G_{\text{sim.}}$) receiving the specified spaces' contents and attributes to reconstruct images. If the structure and the style are correctly decoupled for both experimental and simulated images, the generated simulation image should adopt the same atomic configuration as the original experimental image, exhibiting only a discrepancy in the visual style. Similar scenarios should apply to the original simulation image and the generated experimental image. However, due to the lack of explicit constraints between the simulation and the experimental domains, the above scenarios cannot always be guaranteed for the basic DRIT model. Therefore, a content-consistent loss function $L^{\text{content}}_{\text{consistent}}$ was designed as follows to ensure structure invariability before and after the image transformation:

$$L^{\text{content}}_{\text{consistent}}(x, y, u, v) = \mathbb{E}_{x,y}[\|u-y\|_1 + \|v-x\|_1] \quad (1)$$

where $x$ and $y$ represent the original simulation and experimental images, while $u$ and $v$ represent the generated simulation and experimental images, respectively. Note that only when Trident strategies are implemented together on the DRIT model can the noise-free simulated images be transformed into high-quality ones with a visual style highly resembling the experimental images and atomic structure unchanged (Supplementary Figs. 6-9).

**Performance of the Trident strategy-enhanced DRIT model**

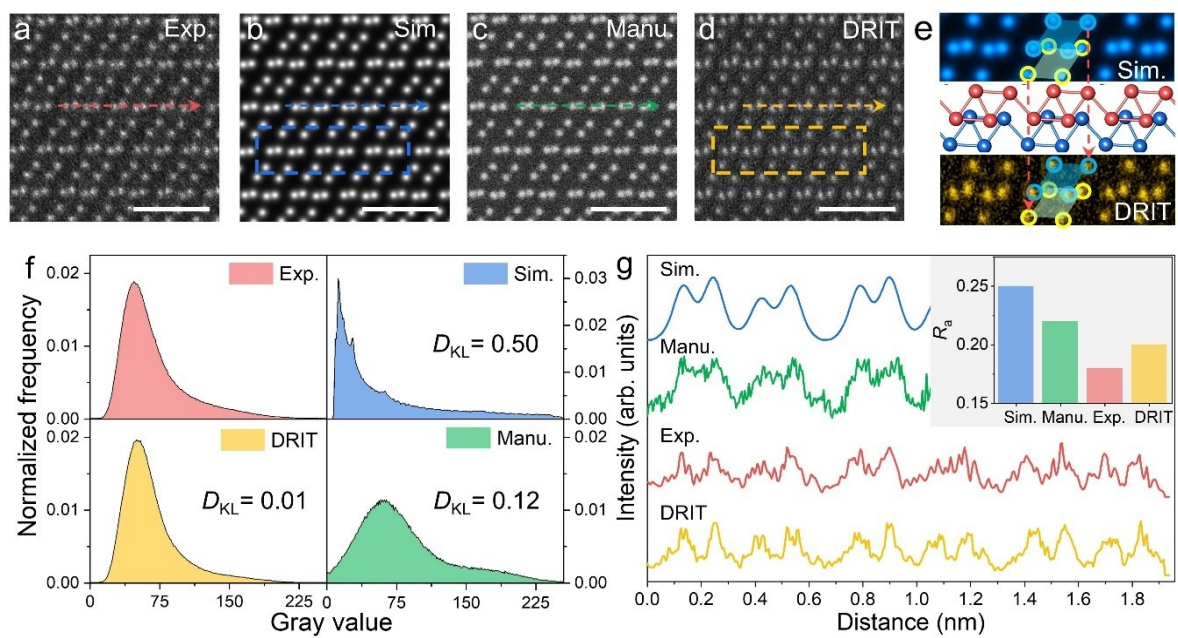

Fig. 2 Performance of the Trident strategy-enhanced generative model. **a**, Experimental annular dark field (ADF) STEM image. **b**, Computem-generated, noise-free simulated image based on the atomic model deduced from (a). **c**, Gaussian noise-added simulated image based on (b). **d**, Disentangled Representation for Image-to-Image Translation (DRIT) model generated simulated image by incorporating structural content in (b) and visual style in (a). Scale bars in (a) to (d): 1 nm. **e**, Zoom-in views of the blue- and orange-boxed regions in (b) and (d), respectively, with the atomic model in between, demonstrating good structural maintenance of the DRIT model. **f**, Grayscale distribution of panels (a) to (d) with their differences quantified by the Kullback-Leibler divergence ($D_{KL}$). **g**, Intensity line profiles taken along the dashed lines in panels (a) to (d), respectively. Inset is the plot comparing the arithmetic mean deviation ($R_a$) of each line profile.

We selected an annular dark field (ADF) STEM image of slip-stacked bilayer $ReS_2$ as a test case (Fig. 2a) and compared the image generation quality from two perspectives: structural consistency and stylistic similarity. Simulation images generated by three different approaches were compared, which are the noise-free simulated image generated by the Computem[47] software based on the manually resolved atomic model corresponding to panel (a) (Fig. 2b), the simulated image constructed by adding Gaussian noise on panel (b) (Fig. 2c), and the Trident strategy-enhanced DRIT model-generated image combining the structure information in panel (b) with the visual style in panel (a) (Fig. 2d). Atom-by-atom comparison between panels (b) and (d) demonstrates good structural retention of the DRIT algorithm without atomic misalignment when conducting style transformation (Fig. 2e). Fig. 2f displays the grayscale distributions of panels (a) to (d), where the DRIT-generated image matches the best with the experimental image regarding the peak location and the full width at half maximum. We introduced Kullback-Leibler divergence ($D_{KL}$) to measure the grayscale distribution difference between the three types of simulated images and the experimental one, showing 0.5 for panel (b), 0.12 for panel (c), and 0.01 for panel (d) (the smaller, the more similar), quantitatively verifying the highest overall stylistic similarity between the DRIT-generated image and the experimental one. The intensity line profiles taken along the same locations in panels (a) to (d) along the dashed lines display local visual style resemblance between the simulated and experimental images (Fig. 2g), where the arithmetic mean deviation ($R_a$) evaluating the curve average fluctuations shows 0.18 for panel (a), 0.25 for panel (b), 0.22 for panel (c), and 0.20 for panel (d) (the closer, the more similar), further supporting the superiority of the DRIT algorithm in style learning. We also investigated the $D_{KL}$ and $R_a$ of simulation images using the JEMs[48] software, yielding 0.36 and 0.25, which is also inferior to the DRIT-generated data (Supplementary Fig. 10). Time complexity analysis shows ∼1 second, ∼2 seconds, and ∼160 seconds cost for generating a 1024×1024 simulation image using the DRIT algorithm, Computem, and JEMs software, respectively, representing an advantage in generation efficiency of our method.

**Structural analysis of the slip-stacked interfaces**

The framework was first applied to resolve the atomic registries of slip-stacked vdW bilayers, which have rotationally aligned top and bottom layers (no interlayer twist) but exhibit sub-angstrom-scale discrepancies in the interlayer sliding, thus showing various physical properties. Although the structural information of different slip stackings is encoded in their complex-valued 2D fast Fourier transform (FFT) (Supplementary Fig. 11) and can be resolved by advanced diffraction techniques like the four-dimensional STEM Bragg interferometry methodology[49,50], the atom-by-atom analysis of the real-space, high-resolution ADF-STEM images is still the most simple and swift means to identify them without high requirements on the equipment. Bilayer $ReS_2$, which was experimentally observed to display diversified slip stacking patterns[15,51], was selected as a test case to evaluate four abilities of this framework: (i) resolving the slip stacking configuration from a raw ADF-STEM image, (ii) quantitatively perceiving subtle structural evolution of the pattern, (iii) accurately localizing the pattern transition interface, and (iv) efficiently performing statistical analysis on large data volumes and contribute to innovative discoveries.

The top panels of Fig. 3a show 6 representative slip stacking patterns involving structures with both overlapping atoms (inconsistent brightness between different atomic columns) and staggered counterparts (uniform brightness of each atomic column). These images were captured from three different shoots having inequable instrument states and aberration parameters. They were input into the structure inference model with neither noise smoothing nor brightness and contrast adjustment (Supplementary Fig. 12a). Our inference model can swiftly figure out the slip vector coordinates, which are subsequently transformed into atomic models automatically (bottom panels) and verified to be correct based on both expert knowledge and image simulation (Supplementary Fig. 12b,c). Note that the accuracy of the model's resolved coordinates depends upon the step size utilized to generate the realistic ADF-STEM simulated image dataset. We used the DRIT-generated bilayer ReS$_2$ images with a step size of 0.05 Å as a test dataset and employed the Euclidean distance $\Delta D$ to evaluate the accuracy of the inferred slip coordinates from different inference models trained by the DRIT-generated images with step sizes ranging from 0.1 to 0.4 Å (in increments of 0.1). The Euclidean distance $\Delta D$ is represented as follows:

$$\Delta D = |\mathbf{D}_{inf} - \mathbf{D}_{gt}| \quad (2)$$

where $\mathbf{D}_{inf}$ is the inferred slip vector from the inference models trained by a generated dataset with a certain step size, and $\mathbf{D}_{gt}$ is the ground truth slip vector from the test dataset with precise labels (inset of Fig. 3b). The boxplot in Fig. 3b exhibits increase of both the average $\Delta D$ and $\Delta D$ corresponding to the middle 95% of the data (box upper limit) as the step size rises, implying degradation of the model accuracy. Considering the balance between the inference model accuracy and the training cost, we chose a step size of 0.1 Å to construct the simulated image dataset, yielding a mean $\Delta D$ of 0.03 Å with 95% of the inferred results deviating less than 0.05 Å from the ground truth, which is suffice for solving our experimental images whose spatial resolution is ∼0.7 Å (Supplementary Figs. 13 and 14).

Our picometer-level accurate framework can be readily generalized to measure faint slip stacking shifts in a large-area STEM image (Fig. 3c). Limited by inferior recognition accuracy and processing throughput of human experts, this image containing more than 3000 atoms was commonly assigned to one slip stacking structure. However, our algorithm took the lower right corner as the reference region and resolved the slip vector nanometer-by-nanometer, revealing ∼0.05 Å/nm slip evolution with subtle interlayer sliding direction variations, as highlighted by the colorful arrows (Supplementary Fig. 15). Such a phenomenon implies that the top and bottom layers are not rigidly and flatly stacked, potentially involving ripples that induce interlayer spacing fluctuations.

The grain boundary often triggers property mutation and is pivotal for microstructure analysis. In slip-stacked bilayer ReS$_2$, an instant stacking pattern transition was observed due to grain boundaries in the bottom layer, which altered the growth direction in the top lattice via vdW epitaxy (Fig. 3d, Supplementary Fig. 16)[52]. We applied a strided pattern matching technique (SPMT) with variable step sizes to balance the global search efficiency and the local analysis resolution at the transition interface (see Methods). A sliding window coarsely scanned across bunches of STEM images with a large stride in the first round. Then, the pattern transition interface was extracted automatically, where a small stride comparable with the length of a chemical bond (step size: 1 Å) was employed to locate the pattern transition interface precisely with the boundary effect mitigated and intricate structural features unveiled (Supplementary Fig. 17a-e). The $D_a$ and $D_b$ mappings in Fig. 3e and Supplementary Fig. 17f display a sharp pattern-switching interface, which agrees well with the human expert knowledge (yellow dashed line) (Supplementary Fig. 18).

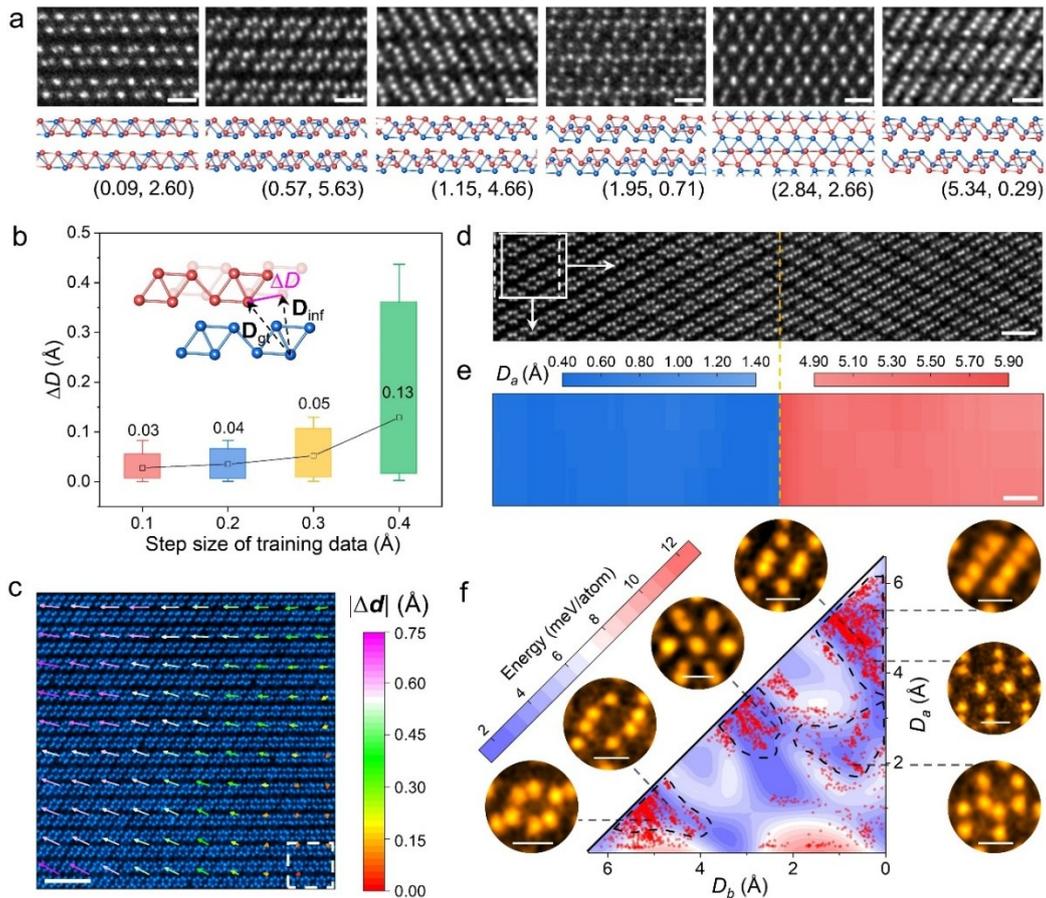

Fig. 3 Automated structural analysis of slip-stacked van der Waals (vdW) bilayers. a, ADF-STEM experimental images of ReS$_2$ bilayers with diversified slip stacking structures (top panels) and their corresponding slip coordinates and atomic models inferred by the ML algorithm (bottom panels). Scale bars: 0.5 nm. b, Boxplot showing the deviation between the inferred ($D_{inf}$) and the ground truth slip vectors ($D_{gt}$), evaluated by the Euclidean distance $\Delta D$ (inset), when the step size of the DRIT-generated simulation image varies from 0.1 to 0.4 Å. c, Spatial variation of slip vectors across a large-area bilayer ReS$_2$. The arrows represent the slip vector deviations ($\Delta d$) between the arrow-located regions and the reference region (white dashed box). The color of the arrow represents the vector length. Scale bar: 2 nm. d, ADF-STEM experimental image displaying an instant pattern transition at an interface marked by the orange dashed line. e, $D_a$ mapping of (d) inferred by the ML model. A strided pattern matching technique (SPMT) is applied to resolve $D_a$ patch-by-patch with a sliding window shown in (d). Scale bars for (d) and (e): 1 nm. f, Slip coordinate distribution of quantities of bilayer ReS$_2$ samples on its potential energy landscape (PEL). PEL is calculated by the density functional theory (DFT) and represents the potential energy fluctuations of slip-stacked bilayer ReS$_2$ at different ($D_a$, $D_b$). The red spots represent the ($D_a$, $D_b$) of image patches with sides of 2 nm. The black dashed lines mark several red dot gathering areas with their typical structures zoomed in around the PEL. Scale bars: 0.3 nm.

We fed 150 pieces of large-scale images involving $\sim 5 \times 10^5$ atoms to the inference model, which were programmatically divided into small patches with a side length of 2 nm for high spatial resolution analysis. ($D_a$, $D_b$) of 3750 patches were resolved within 4 minutes, showing almost two magnitudes of time cost improvement over humans. The slip coordinates are projected onto a density functional theory (DFT) calculated potential energy landscape (PEL) of bilayer ReS$_2$, representing the interlayer energy undulations when one layer slips over another. Two innovative phenomena were unclosed (Fig. 3f): (i) Bilayer ReS$_2$ exhibits diversified slip stacking structures, whose coordinates almost continuously distribute on the PEL diagram, distinctive from most 2D vdW materials (e.g. graphene and MoS$_2$) adopting very limited energetically favorable slip stacking configurations (commonly less than 3 types)[53–56]. (ii) The slip coordinates primarily aggregate in the low energy regions on the PEL, as marked by the black dashed lines, suggesting a thermodynamically driven formation mechanism. The discovery of such a superlubricity-like stacking behavior deepens our comprehension of the vdW epitaxy. It implies a potential approach to constructing diversified slip stackings by direct synthesis, whose key point may lie in selecting low-symmetry vdW materials with weak interlayer coupling, like triclinic $ReS_2$, so that their PELs adopt abundant energy minima with gentle energy undulation (Supplementary Fig. 19), making various slip stacking configurations accessible under mild environment thermal disturbance.

**Robustness and generalizability of the inference model**

The inference accuracy of our model remains robust when the experimental images suffer from a certain concentration of defects or a low SNR. This is particularly significant for STEM analysis of 2D materials with atomic thinness since the electron beam (EB) often induces radiation damage[57,58] like vacancies and holes to these fragile membranes while reducing the EB dosage to suppress the defect generation will inevitably lead to deterioration of the image quality (Fig. 4a). Data augmentations were employed during the inference model training such as adding random proportions of defects, masks, contaminations, and Gaussian noise to the training set (Supplementary Fig. 20). We then used the DRIT-generated realistic STEM simulated images with known structural labels and randomly embedded vacancies to test the inference model's sensitivity to defects and image SNR (Supplementary Figs. 21 and 22, Supplementary Tables 1 and 2). Fig. 4b shows that even if the defect concentration rises to 10% (insets of Fig. 4b), the average $\Delta D$ is still constrained at 0.32 Å with 75% of the total test data yielding $\Delta D$ of less than 0.2 Å (Supplementary Fig. 23a). The model also displays insensitivity to the SNR degradation, where the average $\Delta D$ is stable at ~0.04 Å for the image peak signal-to-noise ratios (PSNR) ranging from 30 to 7 (Fig. 4c). Even when the PSNR decreases to 7 (right inset of Fig. 4c), there still exists more than 90% of the test data having $\Delta D$ of less than 0.2 Å (Supplementary Fig. 23b). Our model also remains robust when the specimens suffer from surface contamination (Supplementary Fig. 23c,d).

The framework was subsequently utilized to resolve experimental images of defective slip-stacked bilayer $ReS_2$ (having holes in one layer, as marked by the white box in Fig. 4d). The Da and Db mappings display similar slip coordinates around the hole with those in the defect-free areas (Fig. 4e and Supplementary Fig. 24a,b), while slight discrepancy is ascribed to the hole-induced lattice deformation. Further investigation manifests the model's parsing capability intact, even if the hole accounts for 25% of a patch area (Supplementary Fig. 24c,d). The model can also rationally provide structural outputs of the experimental image of slip-stacked bilayer ReS2 with grain boundaries (Supplementary Fig. 25). Then, we selected 10 sets of paired high and low SNR STEM images. Each image pair was taken over the same region of the 2D material but with a 4-fold difference in EB irradiation dose. We applied the model to analyze the high and low SNR images and compared the inferred slip vectors between them, finding an average ΔD difference of only 0.15 ± 0.08 Å (Fig. 4f). A typical pair of high and low SNR experimental images are shown in Fig. 4g (i and ii), where the atomic model resolved from the low SNR image (iii) by our algorithm is displayed in the top right corner. The noise-free simulated ADF-STEM image (iv) based on the atomic model agrees well with the high SNR experimental image (i), indicating the inference validity of the low SNR counterpart. The intensity line profiles taken along the same regions in panels (i), (ii), and (iv) exhibit similar line shapes, further demonstrating the strong resolving ability of our framework on experimental images of inferior quality.

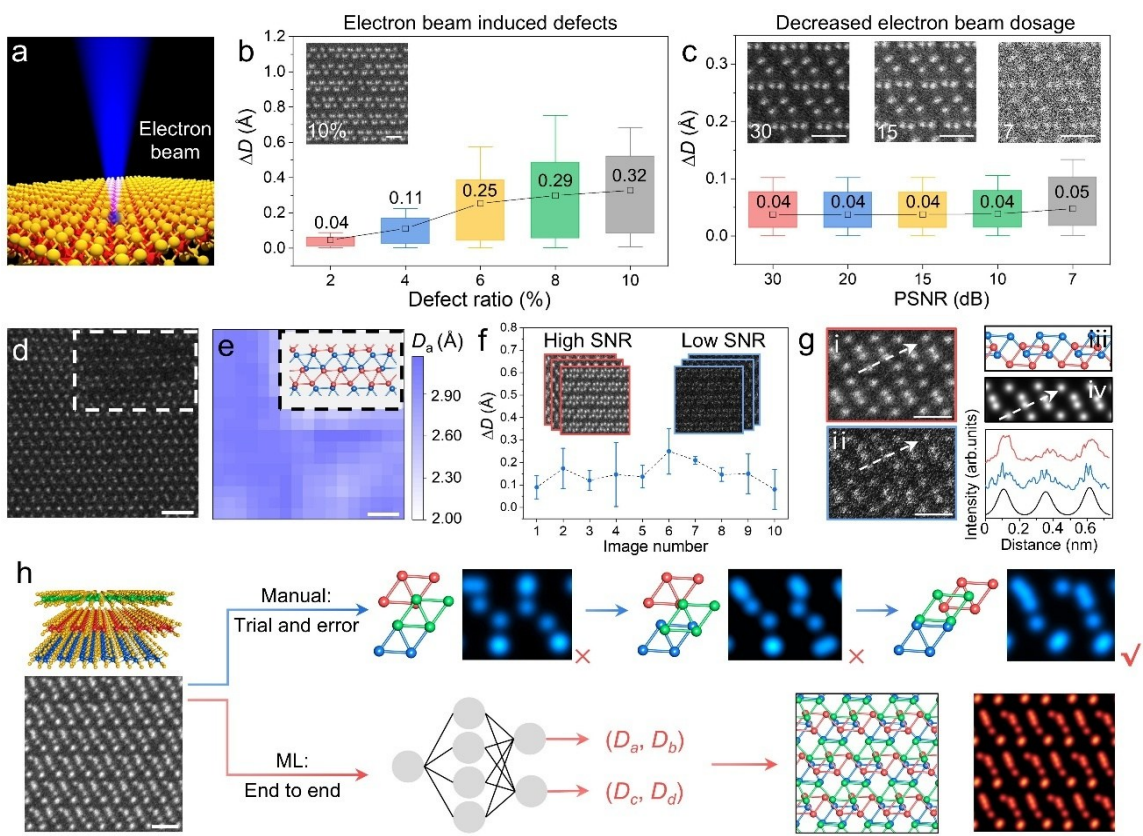

**Fig. 4 Robustness and generalizability of the inference model for slip stacking analysis. a**, Schematics of the electron beam (EB) irradiation on 2D vdW materials. **b**, Inference accuracy as a function of the defect ratio. The inset shows the DRIT-generated image with 10% defect ratio. **c**, Inference accuracy as a function of the peak signal-to-noise ratio (PSNR) of the STEM image. Insets display DRIT-generated images with different PSNR levels. **d**, Experimental ADF-STEM image of a slip-stacked bilayer $ReS_2$ having a hole in one layer (white dashed box). **e**, $D_a$ mapping of (d) inferred by the ML model. Inset is the atomic model of the bilayer region reconstructed by the inferred slip coordinate. **f**, Difference of the inferred slip vectors between experimental images with low and high signal-to-noise ratios (SNR). 10 pairs of experimental images are chosen (insets). **g**, Example verifying the model's reliability when analyzing low SNR experimental images. A pair of experimental STEM images in (f) is displayed in the left panels. The atomic model of the low SNR image is inferred with the corresponding noise-free simulated image shown in the top right corner. Three intensity line profiles taken along the same locations (white arrows) in the high SNR experimental image, the low SNR experimental image, and the noise-free simulated image are shown in the lower right corner. **h**, Generalizability of the ML model to the structural inference of slip-stacked $ReS_2$ trilayers, where the superiority of the end-to-end ML algorithm over the trial-and-error model of human inference is displayed. An experimental ADF-STEM image of a trilayer $ReS_2$ is shown on the left. Scale bars in (b), (c), (g), and (h): 0.5 nm. Scale bars in (d) and (e): 1 nm.

The model can be handily generalized to the structural analysis of slip-stacked trilayers, in which case human experts can only employ trial and error, like playing jigsaw puzzles, to infer potential answers due to structural complexity explosion. The top right panels in Fig. 4h illustrate the reasoning process of humans in the face of a trilayer ReS$_2$ experimental STEM image, where different potential atomic registries between the three layers are sought sequentially with the image simulation conducted one by one until the simulated image matches with the experimental observation. However, our ML approach only requires an extension of the slip stacking structural descriptor from one set of slip coordinates ($D_a$, $D_b$) for the bilayer scenario to two sets of slip coordinates, ($D_a$, $D_b$) and ($D_c$, $D_d$), to represent the interlayer displacements at the 1$^{st}$-2$^{nd}$ interface and the 2$^{nd}$-3$^{rd}$ interface, respectively. Then a structural inference model for trilayer stacks was trained utilizing the DRIT-generated high-quality simulated images with an increased dataset volume, which can readily export atomic registries of trilayer slip stackings in an end-to-end manner (bottom panel in Fig. 4h, Supplementary Fig. 26). In addition, our workflow can also be extended to unclose slip stacking structures of other vdW materials with different crystal structures and elements (Supplementary Figs. 27-30).

**Structural analysis of the twist-stacked interfaces**

The ML framework can also directly resolve the twist angle of vdW materials based on the moiré pattern captured by the STEM image, which is crucial for comprehending the structure-property relationship of such superlattices. One may argue that the interlayer twist can be swiftly figured out from the frequency space by conducting the FFT of the STEM image. However, in some scenarios, such a strategy may have limitations in feasibility and accuracy. First, some moiré patterns corresponding to different twist angles display similar FFT diagrams with indistinguishable disparity. Second, the FFT diagrams of small-area moiré patterns exhibit poor resolution, leading to remarkable errors in angle measurement. All the above circumstances require a method that can directly, accurately, and uniquely assign the twist angle based on the stacking pattern in the real space.

To realize this goal, we conducted modifications to the framework. Distinct from the slip stackings where one set of slip coordinates corresponds to one stacking pattern with a short periodicity length at the sub-nanometer scale, the twist-stacked vdW materials commonly generate superlattices with a much longer periodicity length, which is determined by the twist angle, crystal symmetry, and the detailed lattice parameters. Taking bilayer MoS$_2$ as an example, its periodicity length ($L$)$^{59}$ can be expressed as follows:

$$L = \frac{a}{\sqrt{2(1 - \cos\theta)}} \qquad (3)$$

where $\theta$ represents the twist angle and $a$ is the lattice constant of monolayer MoS$_2$ (0.3165 nm). If the twist angle is 1.5°, the periodicity length reaches 12.08 nm. It is known that the structure inference model takes DRIT-generated simulation image patches with the size of ~2 nm as the input and the interlayer structural descriptor as the output for training. However, the periodicity length of the twist-stacked bilayers often surpasses the patch size. In this case, we use a sliding window with a side length of ~2 nm to scan an area with a side length of not less than $L$ on a DRIT-generated STEM simulation image at a step size ($\Delta L$) of 0.2Å. A set of images acquired at different locations of one moiré pattern is thus collected, which can fully describe all the possible patterns corresponding to one twist angle and are used as the training data for one twist angle (Fig. 5a). Multiple sets of DRIT-generated images and their corresponding twist angles can be achieved in this way and are then utilized for inference model training, yielding an average error as small as 0.29° with 97.5% of the total data having an error of less than 1° (Supplementary Fig. 31).

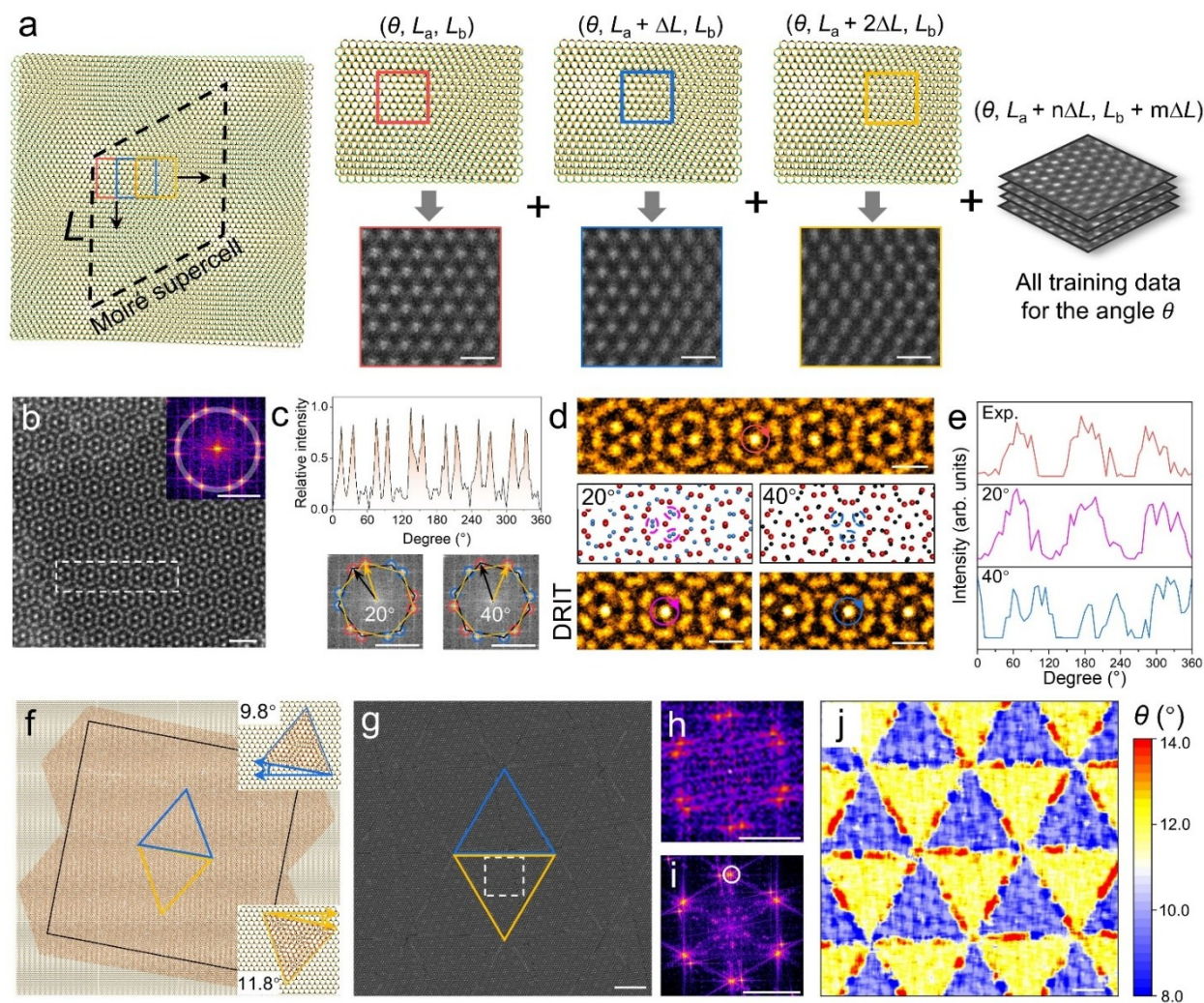

**Fig. 5 Automated structural analysis of twist-stacked vdW bilayers. a**, Schematics illustrating the training dataset acquisition for one twist angle $\theta$. $L$ represents the periodicity length of twist-stacked bilayer $MoS_2$. $\Delta L$ represents the step size of the sliding window to collect image patches for one moiré pattern. $L_a$ and $L_b$ represent the horizontal and vertical coordinates of the initial position of the sliding window (red box). $n$ and $m$ are natural numbers. Scale bars: 0.5 nm. **b**, Experimental ADF-STEM image of twist-stacked bilayer $MoS_2$. Scale bar: 1 nm. Inset is the fast Fourier transform (FFT) of (b). **c**, Infeasibility of uniquely assigning the twist angle of bilayer $MoS_2$ by the FFT pattern. The intensity profile taken along the circle in the FFT of (b) displays similar intensities for 12 reflection spots (top panel), leading to two possible assignments of $\theta$ (lower panels). **d**, Zoom-in view of the white-boxed region in (b) (top panel), atomic models of bilayer $MoS_2$ with twist angles of 20° and 40° (middle panels), and the DRIT-generated images of 20°- and 40°-twisted $MoS_2$ (bottom panels). Scale bars: 0.5 nm. **e**, Intensity line profiles taken along the red, magenta, and blue circles in (d). **f**, Artificially constructed large-area atomic model of bilayer $MoS_2$ where the twist angles of the adjacent triangular areas switch between 9.8° (blue triangle) and 11.8° (yellow triangle). Insets show the magnified views of the atomic structures of twisted stackings with the twist angles of 9.8° and 11.8°. **g**, DRIT-generated image based on the atomic model in the black-boxed region in (f). **h**, FFT of the white-boxed region in (g). **i**, FFT of (g). **j**, $\theta$ mapping of (g) inferred by the ML model. Scale bars in (g) and (j): 5 nm. Scale bars in all FFT diagrams: 4 $nm^{-1}$.

The high-quality simulated data empowers the inference model with the ability to accurately identify the twist angles of highly resembled moiré patterns. Fig. 5b is an experimental ADF-STEM image of twist-stacked bilayer $MoS_2$, whose 12 reflection spots in its FFT pattern corresponding to {100} crystal plane family show similar intensities (top panel in Fig. 5c), leading to two potential twist angle assignments, 20° and 40° (bottom panels in Fig. 5c), if using the FFT pattern to infer. Similar cases will appear in moiré patterns of bilayer MoS$_2$ with twist angles of $\theta$ and 60-$\theta$ ($\theta \in (0°, 60°)$) (Supplementary Fig. 32). Interestingly, our ML algorithm uniquely refers to the twist angle of Fig. 5b as 20°. To judge whether the model inference is correct, we have a closer look at the atomic models of bilayer MoS$_2$ with the twist angles of 20° and 40°, where the separation distances between the adjacent S atomic columns reveal subtle disparity (magenta and blue ovals in the middle panel of Fig. 5d). High-quality simulated images for these two twist angles are then generated via DRIT algorithm (bottom panel in Fig. 5d), in which the intensity profiles taken along the magenta and blue ovals unclose more detailed structural difference in the sulfur peak splitting for the two moiré patterns (Fig. 5e). The intensity feature of the experimental image taken at the same location (red oval in the top panel of Fig. 5d) matches well with that of the 20° twisted simulated image, verifying the credibility of our model's outcome.

Due to spontaneous reconstruction and imperfection in the mechanical transfer, the twist angle, as well as the moiré pattern, may exhibit spatial inhomogeneity[60,61]. To test whether our model can directly resolve the spatial alteration of twist angles in real space, we artificially constructed a large-area atomic model of bilayer MoS$_2$ (Fig. 5f), where the twist angles at different small triangular regions switch between 9.8° (blue triangle) and 11.8° (yellow triangle). Fig. 5g is the DRIT-generated simulation image corresponding to the atomic model in the black-boxed region. If performing FFT to the white-boxed region in which the moiré pattern is homogeneous, the reflection spots are fuzzy due to the small pattern area (Fig. 5h), making accurate measurement of the twist angle infeasible. If FFT is applied to the whole image, the reflection spots marked by the white circles split into two closely located subspots corresponding to 9.8° and 11.8° twist angles without inclusion of any real-space location information (Fig. 5i). Fig. 5j is the twist angle mapping of Fig. 5g via patch-by-patch inference using our model, which agrees with the ground truth well with only imperfections at the domain boundaries due to structural deviation from the intrinsic lattice at these regions. The model's resolving ability to structures involving random spatial variation of the twist angle has also been evaluated, yielding an average inference error within 1° (Supplementary Fig. 33).

**Discussion**

To sum up, we describe a Trident strategy-enhanced DR learning algorithm that solves a key problem in supervised learning, i.e. how to easily achieve training data with high quality and large quantity, which is vital for the application of supervised learning in scientific fields that suffer from scarcity of experimental data, heavy time and labor cost of labeling, and high complexity of the problems. The structural inference model trained by the DRIT-generated high-quality simulated images can directly, rapidly, and accurately figure out atomic-scale configurations at vdW interfaces based on stacking patterns in STEM images across various materials with different stacking modes (slip and twist), layer numbers (bilayers and trilayers), and imaging states (defect ratios, SNR, contaminations) and has the potential to extend to other complicated microstructure analysis. The automated and high-throughput processing ability of the ML method leads to the discovery of a vdW epitaxy mode where diversified thermodynamically favorable slip stackings with almost continuous variations coexist, exhibiting the ML contribution to the knowledge emergence. This work expands the ability of supervised learning from identifying discrete and simple microstructures to analyzing complex and continuously changing motifs. The ML approach demonstrates superiority in efficiency, accuracy, and complexity of problem-solving over human experts, which may revolutionize the characterization and interpretation modes of atomic configurations in microscopy images, paving the way to fast, accurate, automatic, and statistically grounded information extraction of nanomaterials.

## Methods

### Growth and transfer of MX$_2$(M=Re and Mo, X=S) thin films

ReS$_2$ and MoS$_2$ atomically thin layers were grown using hydrogen-free atmospheric pressure chemical vapour deposition (CVD) methods[51,62]. For the growth of ReS$_2$, a space-confined CVD setup was constructed with sodium perrhenate (NaReO$_4$, 99.95%, RHAWN) and sulfur (S, 99.5%, Sigma-Aldrich, 200 mg) used as precursors. 0.01 mol/L NaReO$_4$ aqueous solution was spin-coated onto the c-sapphire substrate surface. Two substrates were placed face-to-face, generating a space-confined reaction microcavity downstream of the furnace. S powder was placed upstream. The rhenium and sulfur sources were maintained at 820°C and 150°C, respectively, for 15 min to produce ReS$_2$ with Ar used as the carrier gas. For the growth of MoS$_2$, molybdenum trioxide (MoO$_3$, 99.5%, Sigma-Aldrich) and sulfur (S, 99.5%, Sigma-Aldrich) powder were used as precursors. To avoid the quench of MoO$_3$ powder by S vapor during the reaction, an inner tube with MoO$_3$ powder placed inside was inserted into the outer 1 in. quartz tube, where S powder was positioned upstream. Two furnaces were applied to give independent temperature control on both two precursors and the substrate. The typical heating temperatures for S, MoO$_3$ and SiO$_2$/Si substrate were ～180, ～300, and ～800 °C, respectively, with a growth time of ～20 min.

For the transfer of MX$_2$ thin layers, a thin film of poly (methyl methacrylate) (PMMA) was initially spin-coated on the MX$_2$/substrate surface. The specimen was then gently floated on a 2 mol/L potassium hydroxide (KOH) solution. When the PMMA/MX$_2$ film was detached from the substrate, the film was transferred to the deionized water three times to thoroughly remove residuals left by the etchant. Next, the film was transferred to a TEM grid, dried naturally in the air, and baked on the hotplate at 180°C for 15 minutes. The PMMA scaffold was finally removed by submerging the TEM grid in acetone for 8 hours.

### STEM characterization

For the slip stacking of bilayer ReS$_2$, ADF-STEM imaging was conducted on an aberration-corrected Titan Cubed Themis G2 300 under an accelerating voltage of 300 kV. Conditions were a condenser lens aperture of 50 mm, convergence semi-angle of 21.3 mrad, and collection angle of 39–200 mrad. The dwell time of a single frame was 2 μs per pixel. A pixel size of 0.012 nm px$^{-1}$ as well as a beam current of 30 pA were used for imaging. For the twisted-stacking of bilayer MoS$_2$, ADF-STEM imaging was conducted using an aberration-corrected JEOL ARM300CF STEM equipped with a JEOL ETA corrector operated at an accelerating voltage of 60 kV located in the Electron Physical Sciences Imaging Centre (ePSIC) at Diamond Light Source. Dwell times of 5−20 μs and a pixel size of 0.006 nm px$^{−1}$ were used for imaging. Optical conditions used a CL aperture of 30 μm, a convergence semiangle of 31.5 mrad, a beam current of 44 pA, and inner−outer acquisition angles of 49.5−198 mrad. Our method demonstrates that ADF-STEM images captured by different types of STEM apparatus with various conditions can be well simulated using a Trident strategy-enhanced DRIT algorithm.

### Noise-free STEM image simulation

Simulated STEM images were generated by the open-source 'incoSTEM' package in Computem. These images can be generated for any material and stacking pattern based on predefined structure files. The structure files were generated in the following four steps using the 'ASE' package: (1) Build a model of the first layer of atoms, consisting of 20 × 20 single cells. (2) Replicate the first layer and adjust the atomic coordinates in the replicated layer by adding displacement in the z-direction, thus achieving the second layer of atoms. (3) Introduce displacement to the replicated second layer along the two in-plane base vector directions, or rotate the replicated layer as a whole around the z-direction in order to generate slip- or twist-stacked bilayer models. (4) Record the coordinates of artificially constructed bilayer models as the structure files, and the slip coordinates ($D_a$, $D_b$) and the twist angles $\theta$ as the labels. These steps were integrated into an automated workflow to efficiently generate simulated images with an average of 1,800 images per hour.

**Data selection**

The high-dimensional STEM image features were first extracted using a pre-trained VGG16 model. Then t-SNE mapped them onto a 2D space by optimizing the position of data points to minimize $D_{KL}$ between high- and low-dimensional distributions. The key t-SNE hyperparameter settings included n-components at 2 and perplexity at 12. K-means clustering was then used to divide the data into K=2 non-overlapping clusters using Euclidean distance in the 2D space of t-SNE reduction. According to the clustering results, the simulation images that are in the same cluster as the experimental images and close to the experimental images were selected as the training data for the DRIT algorithm.

**Trident strategy-enhanced DRIT model training**

The training data of the Trident strategy-enhanced DRIT model consists of unpaired experimental STEM images and noise-free simulated images. The unlabelled raw experimental images utilized for the model training of slip-stacked bilayer $ReS_2$, $ReSe_2$ and twist-stacked $MoS_2$ are all three small photos involving ~ 900 atoms in total, while the noise-free simulated STEM images applied are 41 photos for slip-stacked $ReS_2$, 32 photos for slip-stacked $ReSe_2$ and 119 photos for twist-stacked $MoS_2$. For each dataset, we normalized all the image intensities to [0,1] by their minimum and maximum values and resized them to 1024 × 1024 pixels. Before inputting each patch into training, they were first rotated in the range of -45° to 45°, then randomly cropped to a size from 462 × 462 to 612 × 612 pixels and resized to 1024 × 1024 pixels, and finally flipped (vertically or horizontally) with a probability of 50% for data augmentation. The training procedures used the Adam optimizer with a batch size of 2. All networks of the DRIT model were trained for 3000 epochs, where the learning rate was 1.0 in the first 600 epochs, and linearly decreased to 0 from the 601-3000 epochs. Supplementary Figs. 3 and 8a show detailed model architecture and loss functions.

**Evaluation metrics for simulated image quality**

$D_{KL}$ is a statistical distance of the difference between two probability distributions (P and Q). The more similar the two probability distributions are, the closer their $D_{KL}$ is to 0. It is formulated as follows:

$$D_{KL}(P\|Q) = \sum P \log(\frac{P}{Q}) \tag{4}$$

$R_a$ represents the arithmetic mean of the absolute values of the ordinate $Z(x)$ within the sampling length and reflects the curve average fluctuations. The more similar the two curves are, the closer their $R_a$ values are. It is formulated as follows:

$$R_a = 1/l \int_0^l |Z(x)| \, dx \tag{5}$$

PSNR is a measure of the quality of the noise-adding image in Fig. 4c compared to the original DRIT-generated images, expressed in decibels (dB). Higher PSNR values typically indicate better image quality. It is calculated as follows:

$$PSNR = 10 \times \log_{10}(\frac{MAX^2}{MSE}) \tag{6}$$

where MAX is the maximum possible pixel value and MSE is the mean squared error between the original and the noised image.

**Stacking structure inference model training**

The improved ResNet-50 architecture was used as the backbone of the inference network to achieve the end-to-end quantitative analysis of stacking structures. The full connectivity layer of [1000, 512, $n$] was used as the output header of the model. The size of $n$ was set to 1 for the twisted stacking task, 2 for the slip stacking bilayer task and 4 for the trilayer slip stacking task. Common data augmentation techniques were applied, including random cropping, scaling, flipping, and rotating. To further improve the inference model's robustness to defects, surface contamination, and varying imaging conditions (different doses), other data augmentations were also employed during the inference model training, such as adding random proportions of defects, masks, contaminations, and Gaussian noise to the training set. The training data used here are DRIT-generated simulation images of the pristine lattice with a range of atoms masked and different levels of noise added, which are used to mimic materials having defects, contaminations, and imaged under different doses. The dose variation is realized by adding noise on top of DRIT-generated images. The loss function was designed as follows:

$$\text{Loss} = \lambda_{\text{MSE}} L_{\text{MSE}}(\text{inf, gt}) + \lambda_{L_1} L_{L_1}(\text{inf, gt}) \tag{7}$$

where inf represented the model output, gt represented the ground truth label of the input data, $\lambda_{\text{MSE}}$ was set to 1 and $\lambda_{L_1}$ to 0.5. The training procedures used the Adam optimizer with a batch size of 32. The ResNet-50 was trained for 3000 epochs, where the learning rate linearly increased from 0 to $10^{-3}$ at the first 60 epochs and linearly decreased from $10^{-3}$ to 0 from the 61-3000 epochs.

**Evaluation metrics for structure inference model**

The inference accuracy of the slip stacking structure inference model can be evaluated by $\Delta D$, which measures the Euclidean distance between the slip vector inferred by the model ($\mathbf{D}_{\text{inf}}$) and the ground truth slip vector ($\mathbf{D}_{\text{gt}}$). For a large number of STEM images, we calculate $\Delta D$ for each image and show the overall inference accuracy by the box plots. In Fig. 3b, the central squared spots represent the mean. The color-shaded boxes represent the 5th and 95th percentiles, with the whiskers extending to 3 times the distance between the 5th and 95th percentiles. In Fig. 4b, the boxes indicate the 20th and 80th percentiles with the whiskers extending to 1 time the distance between the 20th and 80th percentiles. In Fig. 4c and Supplementary Fig. 21a,b, the boxes represent the 10th and 90th percentiles, with the whiskers extending to 1.5 times the distance between the 10th and 90th percentiles. In Supplementary Fig. 23c, the boxes represent the 10th and 90th percentiles, with the whiskers extending to 2 times the distance between the 10th and 90th percentiles. In Fig. 4f, the error bar represents the standard deviation of $\Delta D$ measured at three different positions of a pair of high and low SNR images. In addition, the inference accuracy of the twisted stacking structure inference model can be evaluated by $|\Delta \theta|$, which measures the absolute difference between the twist angle inferred by the model ($\theta_{\text{inf}}$) and the ground truth twist angle ($\theta_{\text{gt}}$).

**Strided pattern matching technique (SPMT)**

An ADF-STEM image with thousands of atoms is analyzed by combining the structure inference model with SPMT. As shown in Supplementary Fig. 34, the initial image is first input, and then the STEM image is sampled at a predefined window size and stride length to obtain small patches reflecting local structural features. These patches are analyzed by the inference model to obtain the slip coordinates or twisted angles, and then these outputs are integrated to obtain a 2D mapping, which can show the subtle structural changes and the transition of stacking pattern in the entire image. There are two important parameters in SPMT, the first is the window size, which is set to 2-5 nm. As long as the limit scale of the model is not exceeded, the smaller the sampling size, the finer the local structural features reflected, and the easier to detect the structural variations between different regions. The second is the stride length. The stride length determines the resolution of the local analysis, and the length can be arbitrarily chosen according to the task requirements. As the stride length decreases, the number of patches in the ADF-STEM images and the corresponding total computation time increase exponentially. Therefore, we applied SPMT with two-levels of step sizes to first coarsely locate the region involving pattern transition interfaces and then accurately identify the detailed interface position, thus realizing a good balance between the recognition accuracy and power consumption. As shown in Supplementary Fig. 17, we first chose 1 nm as the initial coarse length to obtain a low-resolution mapping, which is sufficient to identify the interface region of interest. Then, for these regions, the step size is reduced to 1 Å which is close to the length of the chemical bond. Therefore, a high-resolution map of the interface region is obtained, and the interface is accurately located.

**DFT calculations**

DFT calculations were performed using local density approximation (LDA)[63] and implemented in the Vienna Ab Initio Simulation Package (VASP)[64,65]. The cutoff energy of the plane wave is set as 400 eV. The convergence criterion for total energy was set to 1e$^{-5}$ eV and atoms were relaxed until the Hellman-Feynman forces were less than 0.001 eV Å$^{-1}$. The vacuum layer was set to be larger than 10 Å. The k-mesh of 6 × 6 × 1 was adopted for ReS$_2$. Note that, only the vertical coordinates could be relaxed during the sliding.

## Data availability

The experimental and simulation data for Trident strategy-enhanced DRIT model training and the DRIT-generated data for stacking structure inference model training are available on Zenodo (https://doi.org/10.5281/zenodo.11446947).

## Code availability

Codes for DRIT model and stacking structure inference model are available on GitHub (https://github.com/dptech-corp/TED-Gen). Stacking structure inference APP is available on https://bohrium.dp.tech/apps/stacking-pattern-analyzer. The user's manual is described in Supplementary Fig. 35.

## Acknowledgments

S.W. acknowledges support from the National Natural Science Foundation of China (52222201, 52172032), Young Elite Scientist Sponsorship Program by CAST (YESS20200222), Hunan Natural Science Foundation (2022JJ20044), Shenzhen Science and Technology Innovation Commission Project (KQTD202211101115627004), and National University of Defense Technology (ZZCX-ZZGC-01-07). J.Z. acknowledges support from the Ministry of Science and Technology of China (2022YFA1203302, 2022YFA1203304 and 2018YFA0703502), the National Natural Science Foundation of China (Grant Nos. 52021006), the Strategic Priority Research Program of CAS (XDB36030100), the Beijing National Laboratory for Molecular Sciences (BNLMS-CXTD-202001) and the Shenzhen Science and Technology Innovation Commission (KQTD202211101115627004). F.O. acknowledges support from the Key Project of the Natural Science Program of Xinjiang Uygur Autonomous Region(Grant No. 2023D01D03).

## Author contributions

S.W. and J.Z. initiated the project and generated the experimental protocols. W.H., Y.J., Z.G., and L.Y. wrote the code. S.W. prepared the samples and captured the experimental STEM images. J.L. and F.L. conducted DFT calculations. Y.C., Zhemin Li, S.Z., J.C., Zheng Luo, and F.O. discussed the work and gave suggestions. All authors contributed to the data analysis, manuscript writing, and revision of the manuscript.

## Competing interests

The authors declare no competing interests.